\documentclass[preprint,showpacs,prd,amsmath,amssymb]{revtex4}
\usepackage{graphicx}
\usepackage{dcolumn}

\renewcommand{\bar}[1]{\overline{#1}}

\begin{document}

\preprint{USM-TH-127}

\title{ Nuclear Effects on the Extraction of $\sin^2\theta_W$}

\author{Sergey Kovalenko}
\email{Sergey.Kovalenkok@fis.utfsm.cl}
\affiliation{Departamento de F\'\i sica, Universidad
T\'ecnica Federico Santa Mar\'\i a, Casilla 110-V, Valpara\'\i so, Chile}

\author{Ivan Schmidt}
\email{Ivan.Schmidt@fis.utfsm.cl}
\affiliation{Departamento de F\'\i sica, Universidad
T\'ecnica Federico Santa Mar\'\i a, Casilla 110-V, Valpara\'\i so, Chile}

\author{Jian-Jun Yang}
\email{jjyang@fis.utfsm.cl}
\affiliation{Departamento de F\'\i sica, Universidad
T\'ecnica Federico Santa Mar\'\i a, Casilla 110-V, Valpara\'\i so, Chile}
\affiliation{Department of Physics, Nanjing Normal
University, Nanjing 210097, China
}

\begin{abstract}

We study the impact of nuclear effects on the extraction of the
weak-mixing angle $\sin^2\theta_W$ from deep inelastic
(anti-)neutrino-nucleus scattering, with special emphasis on the
recently announced NuTeV Collaboration 3$\sigma$ deviation of
$\sin^2\theta_W$ from its standard model value. We have found that
nuclear effects, which are very important in electromagnetic deep
inelastic scattering (DIS), are quite small in weak charged
current DIS. In neutral current DIS processes, which contain the
weak mixing angle, we predict that these effects play also an
important role and may dramatically affect the value of
$\sin^2\theta_W$ extracted from the experimental data.
\end{abstract}


\maketitle

\section{Introduction}

It is a common belief that the standard model (SM) is a low energy
remnant of some more fundamental theory. In fact, the recent
observation of neutrino oscillations was the first unambiguous
indication of the presence of physics beyond the SM, which allows
for non-zero neutrino masses and lepton flavor violation
processes, forbidden in the SM. This evidence stimulates further
searches for deviations from the SM predictions in physical
observables related to other sectors of the SM.

The precise determination of the weak-mixing angle
$\sin^2\theta_W$ plays a crucial role in testing the SM of
electroweak interactions. Its present value is consistent
with all the known electroweak observables~\cite{CERN}.

A recent announcement by the NuTeV Collaboration~\cite{NuTeV} on a
3$\sigma$ deviation of the value of $\sin^2\theta_W$ measured in
deep inelastic neutrino-nucleus (Fe) scattering with respect to
the fit of the SM predictions to other electroweak
measurements~\cite{CERN}, may be the sign of new physics beyond
the SM. This result takes into account various sources of
systematic errors. However, there still remains the question of
whether the reported deviation can be accounted for by SM effects
not properly implemented in the analysis of the experimental data.
In the present paper we examine the role of nuclear effects on the
extraction of $\sin^2\theta_W$, since an iron nuclear target was
actually used in the NuTeV experiment.

The observables measured in this experiment are ratios of neutral
(NC) to charged (CC) current events, related by a sophisticated
Monte Carlo simulation to $\sin^2\theta_W$. In order to examine
the possible impact of nuclear corrections on the extraction of
$\sin^2\theta_W$, we study the corresponding ratios

\begin{equation}
R^\nu_A = \frac{\sigma (\nu_\mu + A \to \nu_\mu + X)}
{\sigma (\nu_\mu + A \to \mu^- + X)} \label{Rnu},
\end{equation}

\begin{equation}
R^{\bar{\nu}}_A = \frac{\sigma (\bar{\nu}_\mu + A \to \bar{\nu}_\mu + X)}
{\sigma (\bar{\nu}_\mu + A \to \mu^+ + X)} \label{Rnubar}
\end{equation}
of neutral current (NC) to charged current (CC) neutrino
(anti-neutrino) cross sections for a nuclear target A. As is
known, neglecting nuclear effects for an isoscalar target, one can
extract the weak-mixing angle by using the Llewellyn-Smith
relation~\cite{LS}:

\begin{eqnarray}
R_N^{\ \nu(\bar{\nu})} = \frac{\sigma (\nu_\mu(\bar\nu_{\mu}) + N
\to \nu_\mu(\bar\nu_{\mu}) + X)} {\sigma (\nu_\mu(\bar\nu_{\mu}) +
N \to \mu^-(\mu^+) + X)} = \rho^2 \left(\frac{1}{2}
-\sin^2\theta_W + \frac{5}{9} \sin^4 \theta_W ( 1+
r^{(-1)})\right),
\end{eqnarray}
written in terms of NC and CC (anti-)neutrino-nucleon cross
sections. Here,

\begin{equation}
\label{rho}
\rho = \frac{M_W^2}{\cos^2\theta_W M_Z^2}, \ \ \ \ \
r = \frac{\sigma (\bar\nu_{\mu} + N \to \mu^+ + X)}
{\sigma (\nu_\mu + N \to \mu^- + X)}  \sim \frac{1}{2}.
\end{equation}
However, actual targets such as the iron target of the NuTeV
experiment, are not always isoscalar, having a significant neutron
excess. In addition, nuclear effects including the EMC effect,
nuclear shadowing and Fermi motion corrections are known to be
very important for electromagnetic structure functions. These
nuclear effects may also modify the CC and NC structure functions,
and therefore a detailed study of these effects on the extraction
of the weak-mixing angle is essential. In principle any nuclear
model which can successfully explain the EMC effect in deep
inelastic muon-nucleon scattering can serve for our purposes. For
definiteness, in the present work we use a particular nuclear
re-scaling model~\cite{Li2}.  We make complete estimates of
nuclear effects on the ratios $R^{\nu(\bar{\nu})}$ and on the
resulting values of $\sin^2\theta_W$. In order to reduce the
uncertainties related to sea quarks, Paschos-Wolfenstein
~\cite{PW} suggested to extract $\sin^2\theta_W$ from the
relationship

\begin{eqnarray}
R_N^{^-} = \frac{\sigma (\nu_\mu + N \to \nu_\mu + X) - \sigma
(\bar\nu_{\mu} + N \to \bar\nu_{\mu} + X)} {\sigma (\nu_{\mu} + N
\to \mu^- + X) - \sigma (\bar\nu_{\mu} + N \to \mu^+ + X)} =
\rho^2 \left(\frac{1}{2} - \sin^2\theta_W\right).
\end{eqnarray}
Inspired by the Paschos-Wolfenstein relation, we will also examine
nuclear effects on $\sin^2\theta_W$ by the following observable
for the scattering off a nuclear target $A$,

\begin{eqnarray}
R_A^{^-} = \frac{\sigma (\nu_\mu + A \to \nu_\mu + X) -
\sigma (\bar\nu_{\mu} + A \to \bar\nu_{\mu} + X)}
{\sigma (\nu_{\mu} + A \to \mu^- + X) -
\sigma (\bar\nu_{\mu} + A \to \mu^+ + X)}.
\end{eqnarray}

Below we present a
detailed analysis of nuclear effects starting with a brief summary
of the formalism we will use.

\section{Nucleon structure functions}

In the quark-parton model the nucleon structure functions are
determined in terms of the quark
$u(x,Q^2),d(x,Q^2),s(x,Q^2),c(x,Q^2),b(x,Q^2)$ and gluon
$g(x,Q^2)$ distribution functions, which satisfy the QCD
$Q^2$-evolution equations. Below we only collect expressions for
the relevant proton structure functions. The corresponding neutron
structure functions can be obtained from the proton ones by the
replacements $u(x,Q^2) \longleftrightarrow d(x,Q^2)$,
$\bar{u}(x,Q^2) \longleftrightarrow \bar{d}(x,Q^2)$.

The structure functions (SF) of CC reactions $\nu(\bar{\nu}) N \to
l^-(l^+) X$ are given by

\begin{eqnarray} \label{F1WP}
F_1^{W^+ p } &=& \bar{u}(x) (|V_{ud}|^2 + |V_{us}|^2) +
\bar{u}(\xi_b)|V_{ub}|^2 \theta(x_b-x)\\ \nonumber
&+& d(x)|V_{ud}|^2 + d(\xi_c)|V_{cd}|^2 \theta(x_c-x)\\ \nonumber
&+& s(x)|V_{us}|^2 + s(\xi_c)|V_{cs}|^2\theta(x_c-x)\\ \nonumber
&+& \bar{c}(x)(|V_{cd}|^2 + |V_{cs}|^2)
+ \bar{c}(\xi_b)|V_{cb}|^2\theta(x_b-x)\\ \nonumber
&+& b(x)|V_{ub}|^2 + b(\xi_c)|V_{cb}|^2 \theta(x_c-x),
\end{eqnarray}
here $V_{ij}$ are the Cabibbo-Kobayashi-Maskawa mixing matrix elements.
The variable

$$
   \xi_k =\left \{
      \begin{array}{ll}
      x \left( 1 +\frac{m_k^2}{Q^2} \right), & (k=c, b),\\
      x, & (k=u, d, s),
      \end{array} \right.
$$
and the step functions $\theta(x_c-x), \theta(x_b-x)$ take into
account rescaling due to heavy quark production thresholds.

The structure functions $F_2^{W^+ p }$ and $F_3^{W^+ p }$ are
obtained from  (\ref{F1WP}) by the replacements of the quark
distribution functions $q(x, Q^2)$ indicated in the curly
brackets:

\begin{eqnarray}
\label{F2}
F_2^{W^+ p }(x, Q^2)&=& F_1^{W^+ p }(x, Q^2)
\{q(x, Q^2) \to 2x q(x, Q^2), q(\xi_k, Q^2) \to 2 \xi_k q(\xi_k, Q^2)\}, \\
\label{F3}
F_3^{W^+ p }(x, Q^2) &=& 2 \ F_1^{W^+ p }(x,
Q^2)\{\bar{q}(x, Q^2) \to -\bar{q}(x, Q^2)\}.
\end{eqnarray}
The structure functions of the NC reactions $\nu(\bar\nu) N \to
\nu(\bar\nu) X$ are

\begin{eqnarray}
F_1^{Z p } &=& \frac12 \{ [(g_V^u)^2 + (g_A^u)^2] (u(x) + \bar{u}(x) +
c(x) +\bar{c}(x))\\ \nonumber
&& +  [(g_V^d)^2 + (g_A^d)^2] (d(x) + \bar{d}(x) +
s(x) +\bar{s}(x))\},\\ \nonumber
\end{eqnarray}

\begin{eqnarray}
F_2^{Z p } &=& 2 x F_1^{Z p },
\end{eqnarray}

\begin{eqnarray}
\label{FiNC}
F_3^{Z p } &=& 2 [ g_V^u g_A^u (u(x) - \bar{u}(x) +
c(x) - \bar{c}(x))\\ \nonumber
&+&  g_V^d g_A^d (d(x) - \bar{d}(x) +
s(x) -\bar{s}(x))].
\end{eqnarray}
In the SM the vector and axial-vector quark couplings are given by

$$
g_V^u=\frac12-\frac43 \sin^2\theta_W,
\ \ \ g_V^d=-\frac12+\frac23 \sin^2\theta_W, \
\ \ \ g_A^u=\frac12,\ \ \ \  g_A^d=-\frac{1}{2}.
$$
In our analysis we adopt the CTEQ6 Set-2 parton distribution
functions of Ref.~\cite{CTEQ6}.

\section{Nuclear effect on nucleon structure functions}

Here, we summarize our approach for calculating the structure
functions of a given nuclear target starting from the free nucleon
quark distribution functions discussed in the previous section.

\subsection{Nuclear Parton Distributions in the extended $x-$rescaling Model}

Since the discovery of the EMC effect~\cite{SLAC-EMC} in deep
inelastic muon-nucleus scattering, various models~\cite{EMC-SY,EMC-models}
have been proposed for its explanation. In our present work, we choose one
of the successful EMC models usually referred as the extended
$x-$rescaling model~\cite{Li2}. We extend this model to describe
the CC and NC structure functions in (anti-)neutrino-nucleus deep
inelastic scattering.

Let $K^p_{A(N)} (x,Q^2)=x p_{A(N)}(x,Q^2)$, $p=V$, $S$, $G$ be the
momentum distributions of valence quarks(V), sea quarks(S) and
gluon (G) in the nucleus A (or nucleon N), respectively.

The $x-$rescaling model~\cite{Li2} is based on the fact that in a
nucleus the Bjorken variable $x$ of the nucleon structure
functions is rescaled due to the binding energy of nucleons in the
nuclear environment. In Ref.~\cite{Li2} it was found that the
universal $x-$rescaling violates conservation of nuclear momentum.
In the extended $x-$rescaling model this problem is fixed by
introducing different $x-$rescaling parameters for the momentum
distributions of valence quarks and sea quarks (gluons)in the
nucleon structure function, i.e.,

\begin{equation}
 K^{V(S)}_{A} (x,Q^2)=K^{V(S)}_{N} (\delta_{V(S)}x,Q^2).
\end{equation}
For the $x-$rescaling parameters of valence quarks, sea quarks and
gluons we take the values $\delta_V=1.026$ and $\delta_S= \delta_G
= 0.945$~\cite{Li2}. Actually, only one of these two parameters is
free while the other can be determined from the momentum
conservation sum rule.

\subsection{Nuclear shadowing of sea quark and gluon distributions}

The x-rescaling with the above given parameters explains the EMC
effect in the region of medium values of $x$. As is well known, in
the low $x$ region the effect of nuclear shadowing must also be
taken into account. This amounts to a depletion of the nuclear
structure functions at low $x$, due to the destructive
interference of diffractive channels induced by final state
interactions~\cite{nucl-shadow}. In this picture, shadowing
corrections to structure functions are the same for neutrino or
for charged lepton scattering. At low $Q^2$ we expect other
contributions, such as those present in a vector-meson-dominance
model, and in principle in this case there could be differences
between the shadowing corrections to structure functions in
neutrino and in charged lepton scattering. Nevertheless, a
detailed analysis with all the important contributions taken into
account has been performed~\cite{Boros}, with the conclusion that
the total shadowing in neutrino induced reactions is comparable in
magnitude to shadowing in charged lepton induced reactions.
Therefore for our purposes it is enough to take the shadowing
corrections used for nuclear structure functions in lepton charged
scattering, and apply them to the nuclear structure functions of
neutrino induced reactions. For definiteness we use the approach
presented in Ref.~\cite{RSH}, in which this effect is incorporated
into the structure functions by introducing a nuclear shadowing
factor in the sea quark $q_s$ and gluon $g$ distribution functions
$$ q_s(x, Q^2)  \longrightarrow R^A_{sh}(x)\cdot q_s(x, Q^2) , \ \
\  g(x, Q^2)\longrightarrow  R^A_{sh}(x)\cdot g(x, Q^2). $$ Here
we assumed that the gluon and the sea quark distribution functions
receive the same nuclear shadowing. For the nuclear shadowing
factor we use the parameterization proposed in Ref.~\cite{RSH}

\begin{eqnarray}
\label{parameteriz}
R^A_{sh}(x)=
\left\{
      \begin{array}{ll}
      1+a\ln A \ln (x/0.15), & (x < 0.15),\\
      1+b\ln A \ln (x/0.15) \ln (x/0.3), & (0.15 \leq x \leq 0.3)\\
      1, & (x > 0.3).
      \end{array}
\right.
\end{eqnarray}
This parameterization gives the summary of those features of
nuclear shadowing which were important for explaining this nuclear
effect. The parameters a, b in Eq. (\ref{parameteriz}) can be
determined from the experimental data on nuclear shadowing in
$^{56}\rm{Fe}$ and $^{40}\rm{Ca}$~\cite{SLAC-EMC,
Exp-Data-Shadow}. We find the values $a=0.013$ and $b= -0.02$.

\subsection{Nuclear structure functions and Fermi motion of nucleons}

We introduce the average nucleon structure
functions in a nucleus $A$ in a conventional way as~\cite{Li2}

\begin{equation}
\label{NuclSF}
F_i^{(A)}(x,Q^2)=\frac{1}{A} [F_{iA}(x,Q^2)-
\frac 12(N-Z)(F_i^n(x,Q^2)-F_i^p(x,Q^2))],
\end{equation}
where $i=1,2,3$, and $A = Z + N$ with $N$ and $Z$ being the number
of neutron and proton in the nucleus $A$. The functions
$F_{iA}(x,Q^2)$ are the nuclear structure functions. The second
term compensates for the neutron excess in a nucleus $A$. The
functions $F_{i}^{p}(x,Q^2)$  and $F_{i}^{n}(x,Q^2)$ are free
proton and neutron structure functions, calculated according to
Eqs. (\ref{F1WP})-(\ref{FiNC}). The nuclear structure functions
$F_{iA}(x,Q^2)$, with Fermi motion corrections in a nucleus, can
be written as a convolution~\cite{Li1}

\begin{equation}
\label{Fermi} F_{iA}(x,Q^2)=\sum \limits_\lambda\int \frac
{d^3p}{(2\pi)^3} \left|\psi_{\lambda}(\vec{p})\right|^2 z
F_i^{N(A)}(\frac {x}{z},Q^2).
\end{equation}
Here the bound nucleon structure functions $F_i^{N(A)}$ for the
nucleon $N$ in the single-particle nucleon state with wave
function $\psi_{\lambda}(\vec{p})$ are given by
Eqs.~(\ref{F1WP})-(\ref{FiNC}) with the substitutions introduced
in the previous section, in order to incorporate $x$-rescaling and
nuclear shadowing.

In Eq. (\ref{Fermi}) we use the variable
$z=(p_0+p_3)/m_N,p_0=m_N+{\epsilon}_\lambda$, where
$\epsilon_\lambda$ is the binding energy of a nucleon in the
single-particle state $\lambda$. The  single-particle wave
function $\psi_{\lambda}(\vec{p})$  of the nucleon in momentum
space satisfies the light-cone normalization condition:
\begin{equation}
 \int\frac{d^3p}{(2\pi)^3}\left|\psi_{\lambda}(\vec{p})\right|^2z=1.
\end{equation}
In the following calculation, $\epsilon_\lambda$ and
$\psi_{\lambda}(\vec{p})$ are taken from Ref.~\cite{Li2}.

\begin{figure*}
\includegraphics[width=12cm,height=9cm]{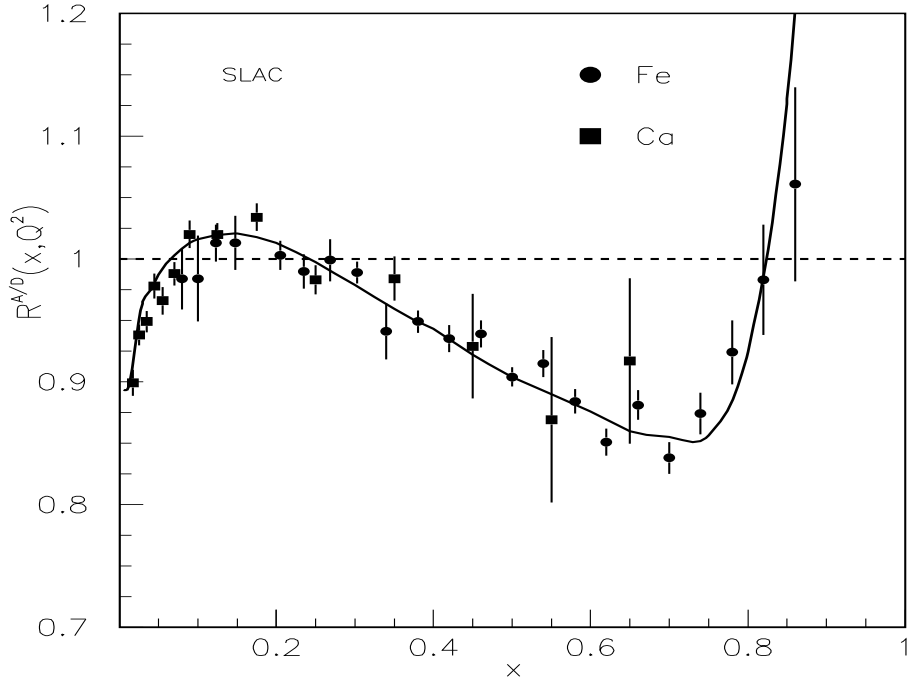}
\caption[*]{\baselineskip
13pt The comparison of our results for
ratio of the nuclear structure function of $^{56}\rm{Fe}$ to that of the
deuteron with the experimental data. $Q^2=20.0~\rm{GeV}^2$ is used.
The experimental data are taken from Ref.~\cite{SLAC-EMC,Exp-Data-Shadow}.}
\label{ksy1f1}
\end{figure*}

Previously, in Ref.~\cite{Li2}, the above described approach was
applied to the analysis of the electromagnetic structure functions
of deep inelastic muon-nucleus scattering in order to explain the
EMC effect~\cite{SLAC-EMC}. In Fig.~\ref{ksy1f1}, we show the
results of this analysis in the form of a ratio of the nuclear
structure functions $F_2^{\gamma}$ of $^{56}$Fe and deuteron. The
experimental points correspond to the data of the EMC
Collaboration~\cite{SLAC-EMC}. As seen from Fig.~\ref{ksy1f1} the
theoretical curve derived in the adopted approach fit the
experimental data with good precision. Fig.~\ref{ksy1f1} also
shows that nuclear effects on the structure functions are
significant.

\begin{figure*}
\includegraphics[width=12cm,height=12cm]{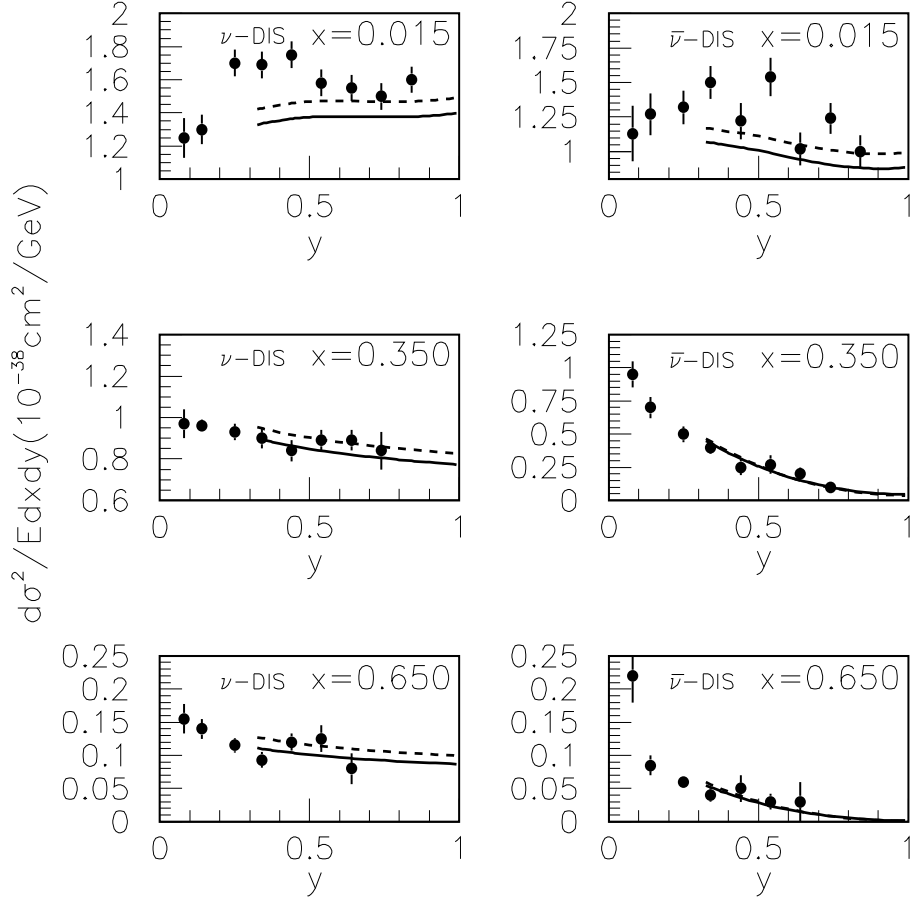}
\caption[*]{\baselineskip
13pt The comparison of our results for
the charged current DIS differential cross sections at
$E_{\nu}=150~\rm{GeV}$ with the experimental data
~\cite{CCFR01}. Our results are given for $y>0.31$ since
the CTEQ6 parton distributions which we used are available only for
 $Q^2>1.3 ~\rm{GeV}^2$. } \label{ksy1f1p}
\end{figure*}

In order to further check our nuclear model, we also apply it to
weak charged current neutrino-iron DIS, which do not depend on the
weak mixing angle. In this case there are no free parameters. The
comparison with data is shown in Fig.~\ref{ksy1f1p}. We see that
the results with nuclear corrections (solid lines) for high and
medium values of $x$ are in excellent agreement with experimental
data~\cite{CCFR01}. At small $x$ the agreement is a bit worse but
still quite reasonable . We also show the results without nuclear
corrections (dashed lines). The surprising conclusion is that
these nuclear corrections are negligible in charged current DIS,
even at large $x \sim 0.65$ values, where in the electromagnetic
case there is a large effect. At small $x$ the data is not very
precise, but the trend indicates that the shadowing corrections
are negligible, and there is even the possibility of
antishadowing. This region certainly deserves further experimental
and theoretical analysis.

Having our approach verified in the case of the electromagnetic
and weak charged processes (structure functions) we apply it to
the analysis of nuclear effects in neutral current
(anti-)neutrino-nucleus deep inelastic scattering, and study their
impact on the extraction of the weak-mixing angle
$\sin^2\theta_W$.

\section{Nuclear effects on the extraction of $\sin^2\theta_W$}

In our numerical analysis we study the influence of nuclear
effects on the extraction of $\sin^2\theta_W$ from the observables
$R_A^{\nu(\bar{\nu})}$ and $R_A^{^-}$, taking into account some
kinematical cut-offs specific for the NuTeV experiment.

The differential cross sections for CC and NC
(anti-)neutrino-nucleus deep inelastic scattering, in terms of the
structure functions defined in Eq. (\ref{NuclSF}), are given by~\cite{Leader}

\begin{eqnarray}
{\frac{d^2 \sigma_{CC}^{\nu, \bar{\nu} }}{dxdy}}^{(A)} &=&
\frac{G_F^2}{\pi}\ m_N\ E_{\nu, \bar\nu}
\left\{ x y^2 F_1^{W^\pm (A)}(x,Q^2) +
\right. \\ \nonumber
 &+& \left. \left(1-y - \frac {xy m_N}{2 \ E_{\nu, \bar\nu}}\right)
F_2^{W^\pm (A)}(x,Q^2) \pm \left(y-\frac{y^2}{2}
\right) x F_3^{W^\pm (A)} (x,Q^2) \right\},
\end{eqnarray}
for the CC reaction, and

\begin{eqnarray}
{\frac{d^2 \sigma_{NC}^{\nu,\bar{\nu} } }{dxdy} }^{(A)} &=&
\frac{G_F^2}{\pi}\ m_N\ E_{\nu, \bar\nu}\left \{ x y ^2 F_1^{Z (A)}(x,Q^2) +
\right. \\  \nonumber
 &+& \left. \left(1-y - \frac {xy m_N}{2 \ E_{\nu, \bar\nu}}\right)
F_2^{Z (A)}(x,Q^2) \pm \left(y-\frac{y^2}{2} \right) x F_3^{Z (A)} (x,Q^2)
\right \},
\end{eqnarray}
for the NC reaction.

In the event selection, the NuTeV Collaboration applied the cut off
\begin{eqnarray}
\label{Ecal} 20 \rm{GeV} \leq E_{cal} \leq 180 \rm{GeV},
\end{eqnarray}
for a visible energy deposit to the calorimeter $E_{cal}$. The
lower limit ensures full efficiency of the trigger, allows for an
accurate vertex determination and reduces cosmic ray background.

Therefore we calculate the observables $R_A^{\nu(\bar{\nu})}$ and
$R_A^{^-}$ imposing the same cut off on the energy $E_h$ of the
final hadronic state $X$, assuming $E_h = E_{cal}$. Since $E_h
\approx \nu$ we can write the kinematical variables averaged over
the (anti-)neutrino flux as

\begin{equation}
x=\frac{Q^2} {2 M_N E_{cal}} \leq 1,\ \ \ \ \ \
y=\frac {E_{cal}} {\langle E_{\nu(\bar{\nu})}\rangle} \leq 1.
\end{equation}
For the average energies of the neutrino and antineutrino beams we
take the values $<E_\nu> = 120~\rm{GeV}$ and
$<E_{\bar{\nu}}>=112~\rm{GeV}$, as in the NuTeV
experiment~\cite{Zeller-private}.

The cut off, Eq. (\ref{Ecal}), characteristic for the NuTeV
experimental events, does not exclude the region of small values
of $Q^2$ where the QCD parton picture is not really applicable.
Therefore it is not possible to calculate the total cross sections
of the CC and NC reactions for the ratios $R_A^{\nu(\bar{\nu})}$
and $R_A^{^-}$ in a theoretically controllable way. Given that we
study these ratios at fixed values of $Q^2$,

\begin{equation}
Q^2=2 m_N \langle E_{\nu(\bar{\nu})}\rangle x y,
\end{equation}
which allows us to examine the $Q^2$ dependence of the nuclear
effects. This method may be helpful if the experimental events
concentrate around some known average value of $\langle
Q^2\rangle$. In the NuTeV experiment $\langle Q^2\rangle \sim 20
\rm{GeV}^2$, but this value was obtained from the Monte Carlo
event simulation. The actual kinematics of the selected CC and NC
events is poorly known, except for the above mentioned energy
deposit cut off (\ref{Ecal}). Despite the fact that the average
$Q^2$ is $20~ \rm{GeV}^2$, a substantial fraction of events may
correspond to relatively low values of $Q^2$. Therefore our
results are not directly applicable to the NuTeV experimental
data, but only indicate some general features of nuclear effects
on the extraction of $\sin^2\theta_W$ from $R_A^{\nu(\bar{\nu})}$
and $R_A^{^-}$ relevant to this experiment, namely that in general
they are sizable. The actual role of nuclear effects can only be
revealed by their inclusion in the corresponding Monte Carlo event
simulation.

\begin{table}
\caption{ Nuclear effects on $R^\nu$ and $R^{\bar{\nu}}$, and
$\delta \sin^2\theta_W$ extracted from $R^\nu$}
\begin{tabular}{|c||c|c|c|}\hline
$Q^2$ &  $~~~\delta R^\nu~~~$ & $~~~\delta R^{\bar{\nu}}~~~$ & $~~~\delta
\sin^2\theta_W~~~$ \\ \hline
 ~~~~3.0~~~~ &
 ~~ -0.001805~~ & ~~-0.000651~~ & ~~ -0.00301~~ \\ \hline
 ~~~~6.0~~~~ &
 ~~-0.000646~~ & ~~ -0.000659~~ & ~~ -0.001050~~ \\ \hline
 ~~~~9.0~~~~ &
 ~~ -0.000018~~ & ~~-0.001081~~ & ~~-0.000018~~ \\ \hline
 ~~~~12.0~~~~ &
 ~~ 0.000328~~ & ~~-0.001811~~ & ~~0.000600~~ \\ \hline
 ~~~~16.0~~~~ &
 ~~0.000639~~ & ~~ -0.002787~~ & ~~0.000960~~ \\ \hline
 ~~~18.0~~~~ &
 ~~0.000747~~ & ~~-0.003195~~ & ~~ 0.001140~~ \\ \hline
  ~~~~24.0~~~~ &
 ~~ 0.000995~~ & ~~-0.004460~~ & ~~ 0.001488~~ \\ \hline
 ~~~~30.0~~~~ &
 ~~ 0.001092~~ & ~~-0.005585~~ & ~~ 0.00164~~ \\ \hline
 \end{tabular}\label{table1}
\end{table}

\begin{table}
\caption{ Nuclear effects on $R^-$, and $\delta \sin^2\theta_W$
extracted from $R^-$}
\begin{tabular}{|c||c|c|}\hline
$Q^2$ &  $~~~\delta R^-~~~$  & $~~~\delta
\sin^2\theta_W~~~$ \\ \hline
 ~~~~3.0~~~~ &
 ~~ -0.000878~~  & ~~ -0.000980~~ \\ \hline
 ~~~~6.0~~~~ &
 ~~-0.000165~~ & ~~ -0.000100~~ \\ \hline
 ~~~~9.0~~~~ &
 ~~ 0.000198~~ & ~~0.000210~~ \\ \hline
 ~~~~12.0~~~~ &
 ~~ 0.000465~~ & ~~0.000509~~ \\ \hline
 ~~~~16.0~~~~ &
 ~~0.000792~~ & ~~0.000840~~ \\ \hline
 ~~~18.0~~~~ &
 ~~0.000912~~  & ~~ 0.000950~~ \\ \hline
  ~~~~24.0~~~~ &
 ~~ 0.001346~~ & ~~ 0.001408~~ \\ \hline
 ~~~~30.0~~~~ &
 ~~ 0.001680~~ & ~~ 0.001780~~ \\ \hline
 \end{tabular}\label{table2}
\end{table}

\begin{figure*}
\includegraphics[width=12cm,height=9cm]{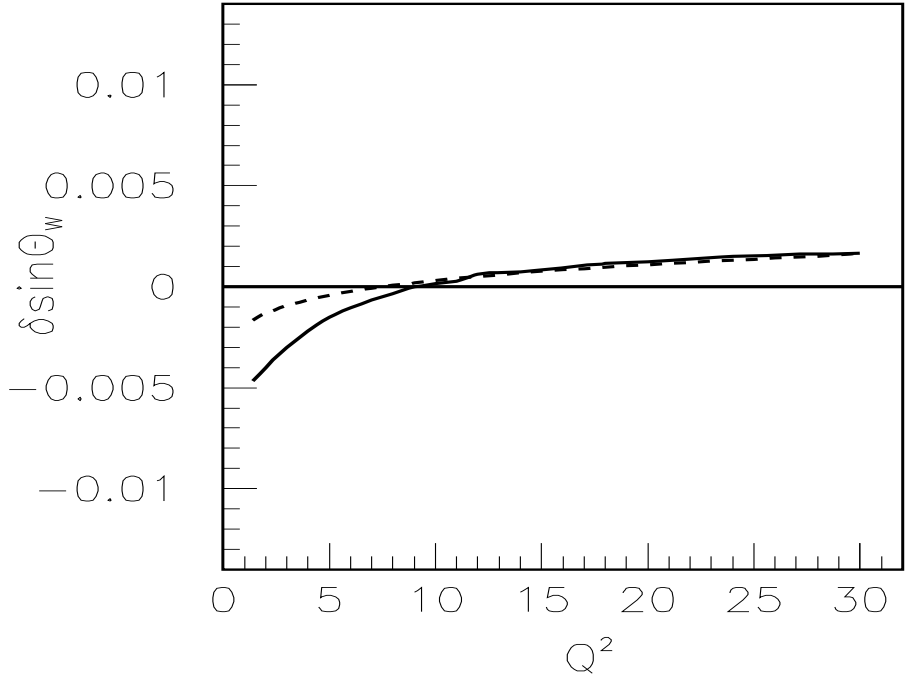}
\caption[*]{\baselineskip
13pt The $Q^2$ dependence of the nuclear modification to
the weak mixing angle $\delta \sin^2\theta_W$ extracted
from $R^\nu$ (solid curve) and from $R^-$ (dashed curve).} \label{ksy1f2}
\end{figure*}

The results of our analysis are summarized in Tables~\ref{table1}
and \ref{table2}, and graphically in Fig.~\ref{ksy1f2}, where we
define $$ \delta R^{i} = R_A^{i} - R_N^{i},
\ \ \ \mbox{with}\ \ \ i = \nu, \bar{\nu}, -. $$ as differences
between the NC/CC ratios calculated with ($R_A^{i}$) and without
($R_N^{i}$) nuclear corrections. The quantity  $\delta
\sin^2\theta_W$ in Tables~\ref{table1} and \ref{table2}, and in
Fig.~\ref{ksy1f2}, are the net effect of nuclear
corrections on $\sin^2\theta_W$ extracted from $R^{\nu}$ (solid curve) or
$R^{^-}$ (dashed curve). The ratio $R^{\bar{\nu}}$  makes no appreciable
influence on  $\sin^2\theta_W$, being weakly sensitive to its
variations in the region of its possible values, and therefore the
corresponding values of $\delta \sin^2\theta_W$ are not presented
in Table 1.

We estimate $\delta \sin^2\theta_W$ in the following way. First,
we calculate the NC/CC ratio $R_N^\nu$ without nuclear effects,
taking for $\sin^2\theta_W$ the central NuTeV value
$\sin^2\theta_W = \sin^2\theta_W^{(N)}=0.2277$ ~\cite{NuTeV}. Then
we calculate the ratio $R_A^\nu$ with nuclear effects in the way
described in section 3, fitting $\sin^2\theta_W$ in order to get
the value of $R_A^\nu$ equal to $R_N^\nu$ calculated in the first
step without nuclear corrections. Thus we obtain the values of
$\delta \sin^2\theta_W$ from the equations
\begin{eqnarray}
\label{deltaS}
&&R_A^\nu(\sin^2\theta_W^{(A)}) = R_N^\nu(\sin^2\theta_W^{(N)}), \\
&& \delta \sin^2\theta_W = \sin^2\theta_W^{(A)} - \sin^2\theta_W^{(N)}.
\end{eqnarray}
The same procedure is applied for the estimation of the nuclear
correction $\delta \sin^2\theta_W$ from the Paschos-Wolfenstein
ratio $R_A^{^-}$.

Therefore in order to get the actual value of $\sin^2\theta_W$,
which is the value $\sin^2\theta_W^{(A)}$ extracted with nuclear
effects, the value $\sin^2\theta_W^{(N)}$, obtained without these
effects, must be shifted by $\delta \sin^2\theta_W$. Note that the
value $\sin^2\theta_W^{(N)}$ corresponds to the case reported by
the NuTeV Collaboration.
In fact, the NuTeV analysis is based on the parton model equations
(7)-(12) for neutrino-nucleon CC and NC DIS, applied to an iron
target nucleus. In this approach one assumes that the parton
distribution functions (PDF) are the effective PDF in iron which
absorb all the nuclear target effects, and are the same for any
DIS process with the same target. The effective PDF in iron were
extracted from CC data~\cite{CCFR01} and then used to calculate
the NC structure functions. However as it follows from our
analysis the PDF per nucleon extracted from the CC data should be
very close to those of a free nucleon, since nuclear effects in CC
DIS are negligible. Thus the analysis based on on the so extracted
PDF deals with practically free nucleon PDF risking to lose the
nuclear effects which do not manifest themselves in the CC DIS but
can be important in the NC neutrino-iron DIS process. Recall that
these effects are very important in charged lepton DIS processes.

 From Tables~\ref{table1} and \ref{table2}, and Fig.~\ref{ksy1f2}, it is
seen that nuclear effects on $R^\nu$, $R^{\bar{\nu}}$, $R^{^-}$
and $\sin^2\theta_W$ are large and strongly depend on the value of
$Q^2$. Moreover, at certain values of $Q^2$ the nuclear correction
$\delta \sin^2\theta_W$  changes its sign. These transition values
are $Q^2\simeq 10 ~\rm{GeV}^2$ and $Q^2 \simeq 8 ~\rm{GeV}^2$ for
the extraction methods using $R_A^{\nu}$ and $R_A^{^-}$
respectively. In the NuTeV experiment the average value of $Q^2$
is about $20 ~\rm{GeV}^2$, which lies in the region of positive
values of $\delta \sin^2\theta_W$. For this reason it might be
thought that nuclear effects enhance the NuTeV deviation of
$\sin^2\theta_W$ from its SM value. However as we already noted, a
substantial fraction of the NuTeV events may have quite small
values of $Q^2$ providing negative nuclear correction $\delta
\sin^2\theta_W$ to the weak-mixing angle $\sin^2\theta_W$.

The following note is also in order. Given the kinematical cut off
in Eq. (\ref{Ecal}), the $Q^2$-region of negative $\delta
\sin^2\theta_W$ corresponds  to the values of the Bjorken variable
$x < 0.25$, where we expect strong nuclear shadowing effects. This
is the case for the parameterization of the shadowing factor given
in Eq. (\ref{parameteriz}). In our analysis shadowing is a
dominant nuclear effect in the region $Q^2 < 5 ~\rm{GeV}^2$
corresponding to $x < 0.13 $. In more sophisticated models
shadowing may be accompanied with antishadowing, acting in such a
way that in the CC DIS their common effect is small due to
self-compensation while in the NC DIS this effect could be larger
than in our model, enhancing the nuclear corrections to the weak
mixing angle $\delta \sin^2\theta_W$ at the small x-values. Thus a
more careful study of these effects in neutrino-nucleus scattering
is required.

Recently, in Ref.~\cite{MT}, it was observed that corrections from
higher-twist effects of nuclear shadowing to $\sin^2\theta_W$
extracted via the Paschos-Wolfenstein relation, may well be of the
same size as the deviation from its global fit value reported by
the NuTeV Collaboration~\cite{NuTeV}.

\section{Summary and Conclusions}

In this work, we have shown that taking into account nuclear
effects such as nuclear $x$-rescaling, nuclear shadowing and Fermi
motion in the nucleon structure functions of CC and NC
(anti-)neutrino-nucleus scattering, may significantly affect the
extracted value of the weak-mixing angle $\sin^2\theta_W$.

The procedure used by the NuTeV Collaboration~\cite{NuTeV} in
order to take into account nuclear effects was to extract
effective nuclear parton distribution functions from charged
current DIS data, which then were used in order to obtain the
value of $\sin^2\theta_W$ from their data. Although in principle
this is a reasonable idea, in practice our results indicate that
it might be difficult to implement, since nuclear effects play a
minor role, if at all, in charged current DIS, while in NC DIS
they are important. Unfortunately there is no direct way of
comparing our theoretical predictions with the results of the
NuTeV Collaboration presented in the form of ratios of NC to CC
experimental event candidates with poorly identified kinematics.
Uncertainties in kinematics are pertinent to the experiments
measuring NC (anti-)neutrino scattering since the final state
neutrinos are not detectable. As a consequence, among the NuTeV
events there might be a substantial fraction of low $Q^2$ events,
despite the fact that in this range their probability decreases
with decreasing $Q^2$. Thus in theoretical estimations of ratios
like $R^\nu$, $R^{\bar{\nu}}$, $R^{-}$, the kinematical
integration in the total NC and CC cross sections extends to the
low $Q^2$ region including $Q^2=0$. This region poses the problem
of calculating the structure functions in a theoretically
controllable way, since the QCD parton model, presently the only
firmly motivated one, is not applicable below $Q^2 \sim$ few
$\rm{GeV}^2$, where QCD has a strong non-perturbative behaviour.

\begin{acknowledgments}
We are grateful to G. P. Zeller and K. S. McFarland  for their
kind correspondence about the NuTeV results~\cite{NuTeV}. We also
would like to thank S. Brodsky for valuable discussions. This work
is partially supported by National Natural Science Foundation of
China under Grant Number 19875024, 10175074; by Foundation for
University Key Teacher by the Ministry of Education (China); by
Fondecyt (Chile) projects 8000017, 1000717 and 3990048, and by
MECESUP FSM9901.

\end{acknowledgments}

\end{document}